\documentclass[twocolumn,aps,showpacs]{revtex4}
\usepackage{graphicx}
\usepackage{bm}
\usepackage{amssymb} 
\begin{document}

\title{Zero-bias tunneling anomaly in a clean 2D electron
gas caused by smooth density variations}

\author{T. A. Sedrakyan, E. G. Mishchenko, and M. E. Raikh}

\affiliation{ Department of Physics, University of Utah, Salt Lake
City, UT 84112}

\begin{abstract}
We show that smooth  variations, $\delta n({\bf r})$, of the local
electron concentration in a clean 2D electron gas give rise to a
zero-bias anomaly in the tunnel density of states, $\nu(\omega)$,
even in the absence of scatterers, and thus, {\em without the
Friedel oscillations}. The energy width, $\omega_0$, of the
anomaly scales with the magnitude, $\delta n$, and characteristic
spatial extent, $D$, of the fluctuations as $(\delta n/D)^{2/3}$,
while the relative magnitude $\delta\nu/\nu$ scales as $(\delta
n/D)$. With increasing $\omega$, the {\em averaged} $\delta\nu$
{\em oscillates} with $\omega$. We demonstrate that the origin of
the anomaly is a weak curving of the classical electron
trajectories due to the smooth inhomogeneity of the gas. This
curving suppresses the corrections to the electron self-energy
which come from the virtual processes involving two electron-hole
pairs.
\end{abstract}


\maketitle

\noindent{\it Introduction.} The origin of a zero-bias anomaly in
the tunnel density of states of disordered metals had been
traced~\cite{79} to the enhancement of the electron-electron
interactions, caused by their diffusive motion. In two dimensions,
the relative correction, $\delta\nu(\omega)/\nu$,
 to the tunnel density of states due to this enhancement is
equal to $(1/2\pi E_{\text {\tiny F}}\tau)\ln(E_{\text {\tiny
F}}^4\tau^3/\omega)\ln(\omega\tau)$~\cite{diffusive}. Here
$\nu=m/\pi\hbar^2$ is the bare density of states, $E_{\text {\tiny
F}}$ is the Fermi energy, $m$ is the electron mass, and $\tau$ is
the scattering time. Diffusive description applies in the energy
domain $\omega\lesssim 1/\tau$. In clean samples with mobility
$\sim 10^6$ cm$^2$/V s  this domain is very narrow, $\sim
10^{-3}$meV. In fact, as it  was demonstrated in
Ref.~\onlinecite{ballisitc}, the 2D zero-bias anomaly extends into
the ballistic regime $\omega \gg 1/\tau$,
and essentially retains its functional form. Virtual processes,
responsible for the anomaly in this regime, involve one impurity
and one electron-electron scattering with either small, $q \ll
k_{\text {\tiny F}}$, or large, $q\approx 2k_{\text {\tiny F}}$,
momentum transfer.

The relative magnitude of the interaction correction,
$\delta\nu/\nu$, falls off with increasing the electron mobility.
As experimental samples become progressively cleaner, the question
arises whether the tunnel density of states in the absence of
impurities exhibits a zero-bias anomaly. This issue was first
addressed in Ref.~\onlinecite{reizer}; the calculation in this
paper predicted the interaction correction of the form
$\delta\nu(\omega)/\nu\propto\omega$.
 However,
later analysis~\cite{antireizer} indicated that, for a finite
interaction range, $d$, the singular behavior, $\delta\nu/\nu =
\omega/4E_{\text {\tiny F}}$, of the correction saturates at
$\omega \lesssim v_{\text {\tiny F}}/d$, where $v_{\text {\tiny
F}}$ is the Fermi velocity.

In the present paper we identify a new mechanism of a zero-bias
anomaly, which is at work for finite-range interactions and {\em
in the absence of impurities}. Namely, we show that a narrow
feature in $\delta\nu(\omega)$ emerges as a result of weak,
long-scale, variations of the electron density, $n({\bf r})$,
which are generic for high-mobility samples. Our main idea is that
the high-order electron-electron scattering processes  in a clean
2D gas, i.e., the processes that involve {\em more than one}
virtual electron-hole pair, are anomalously sensitive to the
variations of $n({\bf r})$. An example of such  process with two
virtual pairs is shown in Fig.~1a. The diagram in Fig.~1a with
three interaction lines describes creation of an electron-hole
pair, which is subsequently rescattered into another pair, and,
finally, annihilated. As was first pointed out in
Ref.~\onlinecite{suhas}, the momenta of states, involved in this
process, are strongly correlated, namely, they are either almost
parallel or almost antiparallel to each other. It is this
correlation that is affected by the spatial inhomogeneity. The
resulting suppression of the contributions of the higher-order
processes, like shown in Fig.~1a, to the self-energy, gives rise
to a zero-bias anomaly. Lack of strong correlation in the momenta
directions in excitation of a {\em single pair} implies that
second-order processes do not contribute to the anomaly.

\begin{figure}[b]
\centerline{\includegraphics[width=75mm,angle=0,clip]{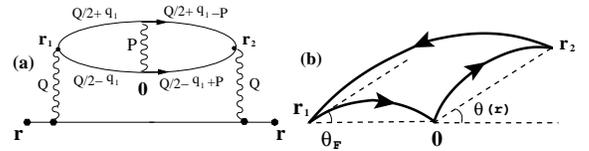}}
\caption{(a) A diagram describing a virtual process of creation,
rescattering and annihilation of the electron-hole pair; (b)
Illustration of lifting the momenta alignment due to curving of
electron trajectories in external field.}
\end{figure}


\noindent{\it Qualitative consideration.} The degree of alignment
of the momenta of states in the diagram  Fig.~1a can be
established from the following consideration. Denote with ${\bf
Q}$ and ${\bf P}$ the momenta transfer in course of creation and
subsequent rescattering of the electron-hole pair. Then the
conditions that the energies
 of all electrons and holes, constituting the pairs, are close
to the Fermi surface can be presented as
$\vert\epsilon_{{\bf q}_1+{\bf Q}/2}-E_{\text {\tiny F}}\vert \sim
\vert\epsilon_{{\bf q}_1-{\bf Q}/2}-E_{\text {\tiny F}}\vert \sim
\omega$, and $\vert\epsilon_{{\bf q}_1-{\bf P}+{\bf Q}/2}-E_{\text
{\tiny F}}\vert \sim \vert\epsilon_{{\bf q}_1+{\bf P}-{\bf
Q}/2}-E_{\text {\tiny F}}\vert \sim \omega$, where $\epsilon_q
=q^2/2m$, and $\omega$ is the energy of the pair. The above
conditions can be met when  either $\vert {\bf Q}\vert$ and $\vert
{\bf P} \vert$ are both small (much smaller than $k_{\text {\tiny
F}}$), or when one of them  is small, while the other is close to
$2k_{\text {\tiny F}}$. For concreteness, we consider the case
$\vert {\bf Q}\vert\approx 2k_{\text {\tiny F}}$, $\vert {\bf P}
\vert \ll k_{\text {\tiny F}}$. Then it follows from the first
condition that $\vert {\bf q}_1 \vert \ll k_{\text {\tiny F}}$,
and that ${\bf q}_1\cdot{\bf Q}\sim \omega k_{\text {\tiny
F}}/v_{\text {\tiny F}}$. Similarly, the second condition requires
that $\left({\bf P}\pm{\bf q}_1\right)\cdot{\bf Q}\sim \omega
k_{\text {\tiny F}}/v_{\text {\tiny F}}$. Combining the two
conditions, we have $\vert\epsilon_{{\bf q}_1+{\bf P}-{\bf
Q}/2}-\epsilon_{{\bf q}_1-{\bf Q}/2} \vert\sim \omega$. The latter
relation can be cast in the form ${\bf P}\cdot\left(2{\bf q}_1 +
{\bf P}-{\bf Q}\right) \sim \omega k_{\text {\tiny
F}}/v_{\text{\tiny F}}$. Since both scalar products ${\bf
P}\cdot{\bf Q}$ and ${\bf P}\cdot{\bf q}_1$ are $\sim \omega
k_{\text {\tiny F}}/v_{\text{\tiny F}}$, we arrive to the estimate
$\vert {\bf P} \vert  \sim \vert {\bf q}_1 \vert \sim k_{\text
{\tiny F}}\left(\omega/E_{\text {\tiny F}}\right)^{1/2}$.
Therefore, the angle between the momenta within the first pair is
small as $\vert {\bf q}_1\vert/k_{\text {\tiny F}} \sim
\left(\omega/E_{\text {\tiny F}}\right)^{1/2}$. Similarly, the
momenta within the second pair are aligned in the angular interval
$\sim \left(\omega/E_{\text {\tiny F}}\right)^{1/2}$.

For the purpose of our derivation, we reformulate the above
restriction in  coordinate space, where $\omega$ defines the
distance, $r$, between the subsequent scattering processes via the
relation $\omega \sim v_{\text {\tiny F}}/r$. Correlation between
the momenta implies that the three  points, ${\bf r}={\bf r}_1$,
${\bf r}= 0$, and  ${\bf r}={\bf r}_2$, in which  creation,
rescattering, and annihilation take place, are located close to
the {\em same} straight line, see Fig. 1b. The ``tolerance'' in
the angle between the vectors ${\bf r}_1$ and ${\bf r}_2$, is the
same as the degree of alignment in the momentum space, $\theta (r)
\sim \left(1/k_{\text {\tiny F}}r\right)^{1/2}$.





In the presence of inhomogeneity, the Fermi momentum, $k_{\text
{\tiny F}}=(2\pi n)^{1/2}$, becomes a function of coordinates. It
is convenient to characterize the random spatial variations of
$n({\bf r})$ by a random force field, ${\bf F}({\bf r})$, related
to the local density gradient as $\nabla n({\bf r})/\langle n
\rangle =e{\bf F}({\bf r})/E_{\text {\tiny F}}$, where $\langle n
\rangle$ is the average density. Denote with $D \gg k_{\text
{\tiny F}}^{-1}$ and $\delta n\ll \langle n \rangle$ the
characteristic scale and the magnitude of the density
fluctuations. Then the typical value of the force is $F\sim
\left(E_{\text {\tiny F}}/eD\right)\left(\delta n/\langle n
\rangle\right)$. The force, ${\bf F}(r)$ curves slightly the
classical electron trajectories transforming them into arcs with
curving angle $\theta_{\text {\tiny \bf F}} =
eF_{\perp}r/2E_{\text {\tiny F}}$, Fig.~1b, where $F_{\perp}$ is
the component of force perpendicular to ${\bf r}$. Obviously, the
process represented by the diagram in Fig.~1a, gets suppressed as
$\theta_{\text {\tiny \bf F}}$ exceeds $\theta (r)$. The condition
$\theta_{\text {\tiny \bf F}}=\theta (r)$ defines the
characteristic distance
\begin{eqnarray}
\label{r0} r_0 \sim k^{-1}_{\text {\tiny F}}(E_{\text {\tiny
F}}k_{\text {\tiny F}}/{eF})^{2/3} ,
\end{eqnarray}
and the corresponding energy scale
\begin{equation}
\label{omega0} \omega_0={v_{\text {\tiny F}}}/{r_0}\sim
{E_{\text {\tiny F}}} {\left(k_{\text {\tiny F}}D\right)^{-2/3}}
\Bigl[{\delta n}/{\langle n \rangle}\Bigr]^{2/3}.
\end{equation}
The latter scale is the energy width of the feature,
$\delta\nu(\omega)$, in the tunnel density of states. As seen from
Eq.~(\ref{omega0}), this scale is determined by the
characteristics of the density variations in combination
$\left(\delta n/D\right)^{2/3}$.


The scales $r_0$ and $\omega_0$ can be derived qualitatively from
a different reasoning. The phase  acquired by the electron upon
travelling the distance $r$, is $\phi(r)=k_{\text {\tiny F}}r$.
Elongation, $\delta{\cal L}$, of the trajectory due to curving,
results in additional phase $\delta \phi(r)=k_{\text {\tiny
F}}\delta {\cal L}=k_{\text {\tiny F}}r(\theta_{\text {\tiny
F}}-\sin \theta_{\text {\tiny \bf F}} )/\theta_{\text {\tiny \bf
F}} \sim k_{\text {\tiny F}} r\theta_{\text {\tiny \bf F}}(r)^2$,
where the curving angle, $\theta_{\text {\tiny \bf F}}(r)$, was
determined above. Curving becomes important when  $\delta
\phi(r)\sim \pi$. This condition yields the same $r=r_0$ as given
by Eq.~(\ref{r0}).

In the above consideration we assumed that the force does not
change within the characteristic distance, $r_0$, between the
collisions. The corresponding condition, $r_0\ll D$, can be cast
in the form
\begin{eqnarray}
\label{condition} \frac{r_0}{D}=\frac{1}{k_{\text {\tiny
F}}D}\Biggl(\frac{E_{\text {\tiny F}}k_{\text {\tiny
F}}}{eF}\Biggr)^{2/3}\!\!\! \sim \;\frac{\left(\langle n \rangle
D^2\right)^{1/2}}{\left(\delta n D^2\right)^{2/3}}\ll 1.
\end{eqnarray}
Eq.~(\ref{condition}) requires that the density variations are
very smooth, $D\gg \langle n \rangle^{3/2}/\left(\delta
n\right)^2$. The other point to be checked upon is whether the
language of the smooth variations of local density, $n({\bf r})$,
that we have  used,  is adequate. Position-dependent  $n({\bf r})$
can be introduced if the {\em statistical} fluctuation,
$\left(\langle n \rangle D^2\right)^{1/2}$,  is smaller than the
change, $\delta n D^2$, of the number of electrons within the
correlation area, $D^2$, due to the {\em smooth} fluctuations. It
is seen from Eq.~(\ref{condition}) that our main condition $r_0
\ll D$ is {\em stronger} than the condition $\delta n D^2 \gg
\left(\langle n \rangle D^2\right)^{1/2}$, so that the reasoning
within the language of local density fluctuations is justified.

It is also instructive to compare the width, $\omega_0$, with
characteristic spatial change of the potential energy of the
electrons, $U=E_{\text {\tiny F}}\delta n/\langle n \rangle$. As
seen from Eq.~(\ref{omega0})
\begin{eqnarray}
\label{comparison} {\omega_0}/{U} \sim (\delta n D^2)^{-1/3}\ll 1,
\end{eqnarray}
so that the anomaly is much
{\em narrower} than the variation of the chemical potential.
Concerning the magnitude, $\delta \nu_0=\delta\nu(\omega_0)$, of
the anomaly, we will establish that
\begin{eqnarray}
\label{magnitude} \frac{\delta\nu_0}{\nu}\sim
\Biggl(\frac{\omega_0}{E_{\text {\tiny F}}}\Biggr)^{3/2}
\!\!\!\sim \;\frac{\delta n D^2}{\left(\langle n \rangle
D^2\right)^{3/2}}\gg \frac{1}{\left(\langle n \rangle
D^2\right)^{3/4}}
\end{eqnarray}
in course of the analytical calculation, to which we turn.
\begin{figure}[t]
\centerline{\includegraphics[width=65mm,angle=0,clip]{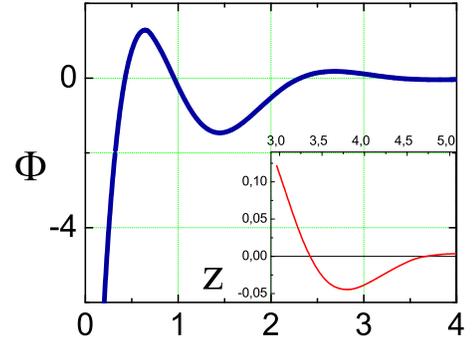}}
\caption{Dimensionless function, $\mbox{\large$\Phi$}(z)$,
describing the shape of the zero-bias anomaly is plotted from
Eq.~(\ref{averaged}) versus dimensionless energy
$z=\omega/\omega_0$. Inset in the lower-right corner:
 enlarged plot of $\mbox{\large$\Phi$}(z)$ in the domain $3<z<5$.}
\end{figure}

\noindent{\it Green functions.} Finding the functional form of
$\delta \nu(\omega)$ amounts, essentially,  to evaluation of the
diagram Fig.~1a in the coordinate space with account of the random
(but locally homogeneous) field, ${\bf F}$. This field enters into
the electron Green function
\begin{equation}
\label{green} G_{\Omega}(0,{\bf r})=\frac{ i^{1/2}~\nu}{\sqrt{2\pi
k_{\text {\tiny F}}r}} \exp\left\{\frac{i\Omega r}{v_{\text {\tiny
F}}} + i k_{\text {\tiny F}}r+i\delta\phi(0,{\bf r})\right\}
\end{equation}
in coordinate-energy space via the additional phase,
\begin{eqnarray}
\label{phi} \delta\phi(0,{\bf r})=\int_0^r k({\bf r}) dl -k_{\text
{\tiny F}}r,
\end{eqnarray}
where $k({\bf r})$ is the wave vector along the classical
trajectory, connecting the points $0$ and ${\bf r}$. Suppose that
${\bf r}$ is directed along the $x$-axis. The parabolic trajectory
is
\begin{eqnarray}
\label{trajectory} y(x)={F_yx(r-x)}/{4E_{\text {\tiny F}}},
\end{eqnarray}
while $dl=dx\sqrt{1+(dy/dx)^2}\approx
dx\sqrt{1+\frac{1}{2}(dy/dx)^2}$.
This allows  to rewrite Eq.~(\ref{phi}) in the form
\begin{equation}
\label{deltaphi} \delta\phi(0,{\bf r})=\frac{k_{\text {\tiny
F}}}{2}\int_0^r\!\!dx\left(\frac{dy}{dx}\right)^2
+\int_0^r\!\!dx\Bigl[k\left\{y(x)\right\}-k_{\text {\tiny
F}}\Bigr].
\end{equation}
Substituting Eq.~(\ref{trajectory}) into Eq.~(\ref{deltaphi}) and
using the relation ${\partial k}/{\partial y}=\bigl[k_{\text
{\tiny F}}F_yy(x)\bigr]/(2E_{\text {\tiny F}})$, we find
\begin{equation}
\label{phase} \delta\phi(0,{\bf r})=-{k_{\text {\tiny
F}}F_y^2r^3}/{96E_{\text {\tiny F}}^2}=-{{k_{\text {\tiny
F}}r\theta_{\text {\tiny F}}^2}}/{24},
\end{equation}
which, within a numerical factor, coincides with the above
qualitative estimate. Naturally, the $x$-component of the field
also contributes to $\delta \phi$. However, this contribution {\em
gauges out} in the expression for $\delta\nu(\omega)$.

\noindent{\it Density of states}. Analytical expression
corresponding to the diagram  Fig.~1a in the coordinate space,
reads
\begin{eqnarray}
\label{deltaG} \delta\nu (\omega)= {\text{Im}}\frac{2i V^3}{\pi^2
\nu^3}\int\frac{d\Omega}{2\pi}\int d{\bf r}\;d{\bf r}_1d{\bf
r}_2\;G_{\Omega}({\bf r},{\bf
r}_1)\qquad\\
\times G_{\omega}({\bf r}_1,{\bf r}_2)\Pi_{\omega-\Omega}({\bf
r}_1,0)\Pi_{\omega-\Omega}(0,{\bf r}_2)G_{\omega}({\bf r}_2,{\bf
r}),\nonumber
\end{eqnarray}
where the polarization operator, $\Pi_{\Omega}({\bf r},{\bf
r}^{\prime})$, is defined  as
\begin{equation}
\label{polar} \Pi_{\Omega}({\bf r},{\bf
r}^{\prime})=-i\int\frac{d\Omega^{\prime}}{2\pi}G_{\Omega^{\prime}}({\bf
r},{\bf r}^{\prime}) G_{\Omega-\Omega^{\prime}}({\bf
r}^{\prime},{\bf r}),
\end{equation}
and $V$ is the dimensionless (multiplied by $\nu$) Fourier
component of the interaction potential, which we assume to be
short-range. We are interested in the oscillatory part of
polarization operator in the presence of the external field.
Substituting Eq.~(\ref{green}) into Eq.~(\ref{polar}), we readily
obtain for this part
\begin{eqnarray}
\label{polar2k} \!\!\!\Pi_{\Omega} (0, {\bf r})\!=\!-\frac{\nu}{2
\pi r^2}\sin\Bigl[2k_{\text {\tiny F}} r-2 \delta\phi({\bf
r})\Bigr] \exp\left\{i\frac{\Omega r}{v_{\text {\tiny
F}}}\right\},
\end{eqnarray}
In Eq.~(\ref{deltaG}) the integration over azimuthal angles of
${\bf r}_1$ and ${\bf r}_2$ can be performed analytically, using
the relation $\langle \exp{i{\bf p}\left({\bf r}_1+ {\bf
r}_2\right)} \rangle_{\varphi_{\bf p},\varphi_{{\bf
r}_1},\varphi_{{\bf r}_2}}= \sin\left[p\left(r_1\pm
r_2\right)+\pi/4\right]/p(r_1r_2)^{1/2}$. Also the integration
over ${\bf r}$ can be carried out explicitly with the help of  the
identity $\int d{\bf r} G_{\Omega}({\bf r}_1,{\bf
r})G_{\Omega}({\bf r},{\bf r}_2)=
\partial G_{\Omega}({\bf r}_1,{\bf r}_2)/\partial \Omega$.
Upon performing these integrations, and combining rapidly
oscillating terms in the integrand of Eq.~(\ref{deltaG}) into
``slow'', oscillating with period $\gg k_{\text {\tiny F}}^{-1}$,
terms, we obtain
\begin{eqnarray}
\label{dos} \delta\nu (\omega)=-\frac{\nu V^3}{2E_{\text {\tiny
F}}\pi^{3/2}k_{\text {\tiny F}}^{1/2}}
\!\int_{r_2>r_1}\frac{d{r_1}d{r_2}}{(r_1r_2)^{3/2}}\int_0^{\omega}\!\!d\Omega\nonumber\\
\times\sin\left[v_{\text {\tiny
F}}^{-1}(\omega-\Omega)(r_1+r_2)\right]
\sum_{\pm}{(r_2\pm r_1)^{1/2}}~~~~~\\
\times\sin\Bigl[{r_1r_2(r_2\pm r_1)}/{r_0^{3}} +{\pi}/{4}\mp
v_{\text {\tiny F}}^{-1}(\omega+\Omega)(r_2\pm r_1)\Bigr]
\nonumber,
\end{eqnarray}
where $r_0=(2^{4/3}/{k_{\text {\tiny F}}})\left({E_{\text {\tiny
F}}k_{\text {\tiny F}}}/{eF_y}\right)^{2/3}$. It is seen that the
characteristic scale of distances $r_1,r_2$ in the integral
Eq.~(\ref{dos}) is indeed equal to $r_0$ in accordance with
qualitative consideration [see Eq.~(\ref{r0})]. The origin of the
combinations $r_1r_2(r_2\pm r_1)/r_0^3$ can be understood from Fig.
1b.
 Scattering sequence
${\bf r}_1 \rightarrow 0 \rightarrow {\bf r}_2 \rightarrow {\bf
r}_1$ leads to the accumulation of the field-dependent phase
$2\left[\delta\phi(r_1)+\delta\phi(r_2)-\delta\phi\left(\vert{\bf
r}_2-{\bf r}_1\vert\right)\right]$. The above combinations emerge
from this additional phase upon using Eq.~(\ref{phase}). Two
contributions to the integral Eq.~(\ref{dos}) correspond to the
locations of the points ${\bf r}_1$ and ${\bf r}_2$ on the
opposite and the same sides from the origin, respectively.

\noindent{\it The shape of the anomaly.} The remaining task is to
perform the gaussian averaging over the random field, $F{\bf}$.
Since $r_0^{-3} \propto F_y^2$, this averaging can be performed
{\em inside the integrand} of Eq.~(\ref{dos}) with the help of the
identity
\begin{eqnarray}
\label{averaging}
\int_{-\infty}^{\infty}\!\!\!dx~e^{-x^2}\!\!\cos(\alpha
x^2+\beta)\!=\! H_1(\alpha)\cos\beta -H_2(\alpha)\sin\beta,\quad
\;\;
\end{eqnarray}
where the functions $H_1$ and $H_2$ are defined as follows
\begin{eqnarray}
\label{functions}
H_{1,2}(\alpha)=({\pi}/{2})^{1/2}\sqrt{(1+\alpha^2)^{-1/2}\pm
(1+\alpha^2)^{-1}}.
\end{eqnarray}
We present the final result in the form $\delta \nu(\omega)/\nu=A
\mbox{\large$\Phi$}(\omega/\omega_0)$, with
\begin{eqnarray}
\label{wm} \omega_0=E_{\text {\tiny F}}\Biggl[\frac{e\langle
F^2\rangle^{1/2}}{\sqrt{2}E_{\text {\tiny F}}k_{\text {\tiny
F}}}\Biggr]^{2/3}\!\!\! =\frac{E_{\text {\tiny
F}}\langle\left(\nabla
n\right)^2\rangle^{1/3}}{\left(4\pi\right)^{1/3}\langle n
\rangle},
\end{eqnarray}
in agreement with qualitative estimate Eq.~(\ref{omega0}), and with
constant $A$  defined as
\begin{equation}
\label{A} A=-\frac{V^3v_{\text {\tiny F}}}{4\pi E_{\text {\tiny
F}}k_{\text {\tiny F}}^{1/2}r_0^{3/2}}
= -\frac{V^3}{8\sqrt{2} \pi^{3/2}} \Biggl[\frac{\langle\left(\nabla
n\right)^2\rangle}{\langle n \rangle ^3}\Biggr]^{1/2};
\end{equation}
the dimensionless function, $\mbox{\large$\Phi$}(z)$, describing
the shape of the anomaly, is given by
\begin{figure}[t]
\centerline{\includegraphics[width=85mm,angle=0,clip]{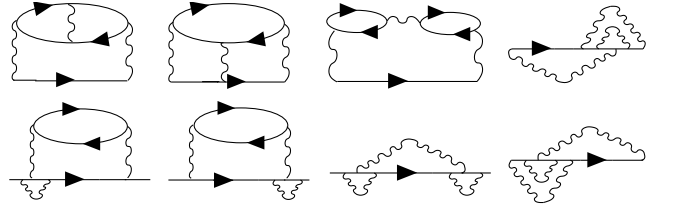}}
\caption{Diagrams representing all third order processes with {\it
aligned} momenta of the virtual states.}
\end{figure}

\begin{eqnarray}
\label{averaged} &&\!\!\!\!\!\mbox{\large $\Phi$}(z)=
\!\int\limits_{\rho_2>\rho_1}\!
\frac{d{\rho_1}d{\rho_2}}{(\rho_1\rho_2)^{3/2}}\!\!\int\limits_0^z\!dz^{\prime}
\sin\Bigl[(z-z^{\prime})(\rho_1+\rho_2)\Bigr]\qquad\nonumber\\
&&\!\!\!\!\!\Bigl\{S_{+}(\rho_1,\rho_2)\!+\!C_{+}(\rho_1,\rho_2)\!+\!S_{-}(\rho_1,\rho_2)\!+
\!C_{-}(\rho_1,\rho_2)\Bigr\}\nonumber\\
&&\!\!\!\!\!=\mbox{\large $\Phi$}_{+}(z)+\mbox{\large
$\Phi$}_{-}(z),
\end{eqnarray}
where the functions $S_{+}$, $S_{-}$, $C_{+}$, and $C_{-}$ are
defined as
\begin{eqnarray}
\label{functions1}
S_{\pm}(\rho_1,\rho_2)={(\rho_1\pm\rho_2)^{1/2}}
\sin\left[\frac{\pi}{4}\mp (z+z^{\prime})(\rho_1\pm \rho_2)\right]\nonumber\\
\times\Biggl\{ H_1\biggl(\rho_1\rho_2(\rho_1 \pm \rho_2)
\biggr)-\sqrt{\pi}\Biggr\},\qquad\qquad\qquad\qquad\\
C_{\pm}(\rho_1,\rho_2)={(\rho_1\pm\rho_2)^{1/2}}
\cos\left[\frac{\pi}{4} \mp (z+z^{\prime})(\rho_1\pm \rho_2)\right]\nonumber\\
\times
H_2\biggl(\rho_1\rho_2(\rho_1\mp\rho_2)\biggr).\qquad\qquad\qquad\qquad\qquad\qquad
\end{eqnarray}
In definitions of $S_{+}$ and $S_{-}$ we had subtracted from the
function $H_1(\alpha)$ the zero-$F$ value $H_1(0)=\sqrt{\pi}$.
Integration over $z^{\prime}$ in Eq.~(\ref{averaged}) can be
carried out analytically. The remaining integrals over $\rho_1$,
$\rho_2$ were evaluated numerically. The resulting shape of the
zero-bias anomaly is shown in Fig.~2. The small-$z\ll 1$ behavior
of  $\mbox{\large$\Phi$}(z)$ is $8\ln z$, i.e. it diverges
logarithmically. The cutoff is chosen from the condition that
$\mbox{\large$\Phi$}(z)$ approaches zero at large $z$. Note, that
$\mbox{\large$\Phi$}(z)$ exhibits a pronounced feature around
$z=1$. The origin of this feature lies in strong oscillations of
the integrand in Eq.~(\ref{dos}). The ``trace'' of these
oscillations {\em survives} after averaging over the magnitude of
the random field. In fact, the oscillations persist beyond $z=3$,
as seen in the inset in Fig.~2. Also analytical inspection of
Eq.~(\ref{averaged}) for $z\gg 1$ yields
\begin{eqnarray}
\label{oscillation} \Phi_{+} (z)\mbox{\Large$|$}_{z\gg
1}\approx-2^{3/4}\sqrt{\pi}\;\frac{\sin\left(2^{8/3}\sqrt{3}z\right)}{z^{3/4}}\exp\left\{-2^{8/3}z\right\}.\;\;
\end{eqnarray}

\noindent{\it Other third-order processes}.
Eqs.~(\ref{wm})-(\ref{averaged}) were derived for a specific
process, illustrated in the diagram in Fig.~1a. However, creation,
rescattering, and annihilation of a pair can follow a different
scenario, {\em e.g.}, rescattering process can involve the initial
electron, as illustrated by the second and third diagrams in
Fig.~3. Important is, that the restriction concerning the momenta
alignment, leading to the zero-bias anomaly, applies to this
scenario as well. It also applies to {\em all} other diagrams in
Fig.~3. Note, that diagrams in Fig.~3 do {\em not} exhaust
possible third-order processes~\cite{suhas}. All contributions to
$\delta \nu$ of the diagrams in Fig.~3 have the same analytical
structure and differ only by numerical coefficients, originating
from  spin degeneracy and from closed fermion loops [each bringing
a factor $(-2)$]. Collecting these contributions, amounts to
multiplying the first diagram in Fig.~3 by $1/2$.


\noindent{\it Concluding remarks}. Higher order processes in a
homogeneous electron gas, involving $n>2$  electron-hole pairs are
also subject to the momenta restriction  \cite{suhas}, leading to
the anomaly in the presence of inhomogeneity. However, these
processes are suppressed as $(\omega_0/E_{\text {\tiny F}})^{n/2}$
due to the phase-space limitation.

Note, that in addition to the oscillating term, the polarization
operator Eq.~(\ref{polar2k}) contains also a slow-varying term
$\propto \vert \Omega \vert/r$. Evaluating $\delta\nu(\omega)$
from Eq.~(\ref{deltaG}) with this slow part of $\Pi_{\Omega}$
yields a non-anomalous correction. The same applies to all
diagrams in Fig.~3. Moreover, summation of RPA subsequence of the
diagrams with ``slow'' $\Pi_{\Omega}({\bf r})$, a procedure
similar to Refs.~\onlinecite{reizer},~\onlinecite{antireizer},
does not produce any anomaly, if the interactions are
short-range~\cite{antireizer}.

In experimental situations, the electrons are supplied to the 2D
gas by donor impurities, separated from electrons by a wide
spacer. Growth-related technological inhomogeneities, like
ridges~\cite{willett}, do not change the average electron density,
but redistribute electrons over the plane~\cite{zhitenev}, and set
scales $\delta n$ and $D$. In principle, {\em individual donors
themselves} create the Friedel oscillations of the electron
density.  However, these oscillations are exponentially suppressed
due to the large separation of donors from the 2D gas. A question
might be asked as to why in evaluating of diagram in Fig. 1a we
had neglected violation of the momentum conservation due to the
impurity scattering. The answer is that condition
Eq.~(\ref{condition}) justifies such a neglecting. This is because
at distances larger than the Bohr radius, the donor potential is
screened. Thus, instead of individual donors, the electron is
scattered by a smooth in-plane potential  with spatial scale, D.
Then the scattering angle does not exceed $1/(k_{\text {\tiny
F}}D)$. On the other hand, the relevant curving angle,
$\theta_{\text {\tiny F}}$,  is $(\omega_0/E_{\text {\tiny
F}})^{1/2}\sim [(\delta n/n)]^{1/3}(k_{\text {\tiny F}}D)^{-1/3}$,
as follows from Eq.~(\ref{omega0}). Condition
Eq.~(\ref{condition}) guarantees that this angle is bigger than
$1/(k_{\text {\tiny F}}D)$. Summarizing, borrowing momentum from
donors is not efficient, since they are distant and screened.

This conclusion is also supported by comparison of the inverse
{\em transport} scattering time, $\tau_{tr}^{-1} \sim E_{\text
{\tiny F}}(U/E_{\text {\tiny F}})^2(1/k_{\text {\tiny F}}D)$, from
screened impurities and the energy scale $\omega_0$. As follows
from Eq.~(\ref{omega0}) the ratio $1/(\omega_0\tau_{tr}) \sim
(U/E_{\text {\tiny F}})^{4/3}(1/k_{\text {\tiny F}}D)^{1/3}$ is
small. The latter also implies that the diffusive anomaly of
Ref.~\onlinecite{diffusive} develops at $\omega$
much smaller than $\omega_0$.

Discussions with  A. Andreev,  A. Chubukov, L. Glazman, B. Halperin,
D. Maslov, and   K. Matveev are gratefully acknowledged. E.M.
acknowledges the support of DOE Award No. DE-FG02-06ER46313.

\end{document}